\documentclass[pra,aps,eqsecnum,twocolumn,showpacs]{revtex4}
\usepackage{mathtext}
\usepackage[T2A]{fontenc}
\usepackage[cp1251]{inputenc}
\usepackage{epsf}
\usepackage{psfrag}
\usepackage{graphicx}
\usepackage{amssymb,amsmath}

\begin{document}

\title{Light trapping in an ensemble of point-like impurity centers in Fabry-Perot cavity}
\author{A. S. Kuraptsev and I. M. Sokolov
\\
{\small Peter the Great St. Petersburg Polytechnic University, 195251, St. Petersburg, Russia}\\
}
\date{\today}

\sloppy



\begin{abstract}
We report the development of quantum microscopic theory of quasi-resonant dipole-dipole interaction in the ensembles of impurity atoms imbedded into transparent dielectric and
located into Fabry-Perot cavity. On the basis of the general approach we study the simultaneous influence of the cavity and resonant dipole-dipole interaction on the shape of
the line of atomic transition as well as on light trapping in dense impurity ensembles. We analyze this influence depending on the size of the ensemble, its density, as well as
on r.m.s. deviation of the transition frequency shifts caused by the symmetry disturbance of the internal fields of the dielectric medium. Obtained results are compared with
the case when the cavity is absent. We show that the cavity can essentially modify cooperative polyatomic effects.
\end{abstract}


\maketitle
\section{Introduction}

Ensembles of impurity centers (atoms) imbedded into dielectric matrix are considered as promising objects for wide range of problems in modern quantum optics and quantum
electronics. The efficiency of their use, especially for optical applications, depends to a large measure on optical depth of the ensembles.  Optical thickness can be increased
by extension of the system or/and by decreasing of the mean free path of photons inside it.

In many cases increasing of the size is undesirable or even impossible. In this situation the simplest and direct way to increase the optical depth is increasing of the density
of impurities. This way enhances the collective effects especially when the photon mean free path becomes comparable or less than resonant wavelength. In such a case resonant
dipole-dipole interatomic interaction leads to density-dependent broadening and shifts of atomic transition as well as distortion of spectral line shape. Collective effects in
the dense atomic ensemble are studied in detail both theoretically, to name a few \cite{Fr}-\cite{17a}, and experimentally for cold atomic gases \cite{22}-\cite{b} (see also
references therein).

Besides resonant interatomic interaction the line shape of impurity centers is transformed due to interaction with the surrounding dielectric medium (see \cite{6} -- \cite{16}).
Even in the case of a transparent dielectric the internal fields of a medium cause spectral line shifts of the impurity atoms which can exceed the natural linewidth. These
shifts depends mainly on the type of chemical bond of a dielectric, the symmetry of the internal fields and the temperature.

Another way to increase efficiency of light interaction with impurity ensembles is to use optical cavity or waveguide. A cavity offers an exciting tool to control over the
light-matter interaction. Since the seminal work of Purcell \cite{1} the peculiarities of atomic radiative properties, in particular, the enhancement and inhibition of the
spontaneous decay rate  inside a cavity or waveguide as well as near its surface has attracted a considerable attention \cite{2} -- \cite{5}. Light matter interface in the
presence of nanophotonic structures, such as nanofibers \cite{5a} -- \cite{5b1}, photonic crystal cavities \cite{5c} and waveguides \cite{5d} -- \cite{5e} propose future
applications for quantum metrology, scalable quantum networks and quantum information science.

Cavity modifies the structure of modes of electromagnetic field. It causes not only modification of spontaneous decay but also the nature of photon exchange between different
atoms. In its turn it leads to alteration in dipole-dipole interatomic interaction \cite{25}. As opposed to spontaneous decay the modification of the dipole-dipole interaction
is studied in less detail. By now there are several works dedicated to the interatomic interaction in the atomic systems coupled to a nanofiber \cite{25a} as well as to phonic
crystals \cite{25b} -- \cite{25d}. Nevertheless, polyatomic cooperative effects inside the cavity, including multiple and recurrent scattering have not  been studied in detail
yet.

The main goal of the present work is to analyze theoretically polyatomic cooperative effects in an ensemble of point-like impurity centers embedded in a solid dielectric into
the Fabry-Perot microcavity. We developed consistent quantum theoretical approach based on approximate calculation of wave function of the joint system consisting of $N\gg1$
motionless centers and the electromagnetic field. The interaction of impurity atoms with the dielectric is simulated by introduction inhomogeneous level shifts of the atomic
energy levels.

As an example of a practical implementation of this approach in the present work we consider spontaneous decay of the local atomic excitation prepared inside a cavity. We
calculate the transition spectrum of an excited atom and study radiation trapping in considered system.  We analyze the role of the cavity depending on the size of impurity
ensemble, its density  as well as on rms deviation of the transition frequency shifts of the impurities caused by internal fields of the dielectric medium. The special attention
is given to the case when the distance between two mirrors is less than a half of the transition wavelength. This case is of particular interest due to practically complete
suppression of spontaneous decay of some Zeeman sublevels of atomic exited state. Obtained results are compared with the case when the cavity is absent.

\section{Basic assumptions and approach}
Let us consider an ensemble, which consists of $N$ motionless impurity atoms imbedded into transparent dielectric and placed into a Fabry-Perot cavity. The mirrors of a cavity
are assumed to be perfectly conducting. The longitudinal sizes of the mirrors are much larger than resonant transition wavelength $\lambda_{0}$, the distance between the mirrors
$d$ and the average distance between impurity atoms. It will allow us to consider the limit of infinite sizes of the mirrors in the final expressions.

We assume that the temperature is low enough to neglect the electron-phonon interaction.  Influence of the dielectric on impurity atoms is simulated by means of random shift of
their energetic levels.  We think that the transition frequency of impurity atoms in a dielectric $\omega_{a}$ differs from the transition frequency of a free atom $\omega_{0}$
-- $\omega_{a}=\omega_{0}+\Delta_{a}$, where $\Delta_{a}$ is the frequency shift of the atom $a$ ($a=1,...,N$) which depends on its spatial position due to inhomogeneity of
internal fields in dielectric.

In this paper we use quantum microscopic approach described firstly in \cite{27} and developed afterward in \cite{21} for description of collective effects in dense and cold
nondegenerate atomic gases. This approach was earlier successfully used for analysis of  optical properties of dense atomic ensembles \cite{18} -- \cite{20} as well as for
studding of light scattering from such ensembles \cite{22a} -- \cite{26}.

It is based on solution of the non-stationary Schrodinger equation for the wave function $\psi$ of a joint system consisting of atoms and the electromagnetic field.
\begin{equation}\label{1}
  i\hbar\frac{\partial\psi}{\partial t}=\widehat{H}\psi.
\end{equation}
The Hamiltonian $\widehat{H}$ of the joint system can be presented as a sum of Hamiltonian $\widehat{H}_{a}$ of the atoms noninteracting with the field, the Hamiltonian
$\widehat{H}_{f}$ of the free field in a Fabry-Perot cavity, and the operator $\widehat{V}$ of its interaction.
\begin{eqnarray}\label{2}
  \widehat{H} &=& \widehat{H}_{0}+\widehat{V}, \\
  \widehat{H}_{0} &=& \widehat{H}_{f}+\sum_{a}\widehat{H}_{a}.
\end{eqnarray}
In the dipole approximation used here, we have
\begin{equation}\label{3}
  \widehat{V}=-\sum_{a}\widehat{\textbf{d}}^{(a)}\widehat{\textbf{E}}(\textbf{r}_{a}).
\end{equation}
In this equation $\widehat{\textbf{d}}^{(a)}$is the dipole momentum operator of the atom \emph{a}, $\widehat{\textbf{E}}(\textbf{r}_{a})$ is the electric field operator, and
$\textbf{r}_{a}$ is the position of the atom $a$.

The electric field operator $\widehat{\textbf{E}}(\textbf{r})$ in a microcavity can be obtained in a standard way by quantization of the classical field
$\textbf{E}(\textbf{r},t)$. The latter is a solution of Maxwell equations with corresponding boundary conditions.

Let us consider coordinate system with $z$ axis perpendicular to the mirrors and with reference point $z=0$  at one mirror. In such a case the boundary conditions can be written
as follows: $E_{x}|_{z=0}=E_{x}|_{z=d}=E_{y}|_{z=0}=E_{y}|_{z=d}=0$. Solving the Maxwell equations, we have
\begin{equation}\label{4a}
  \textbf{E}(\textbf{r},t)= \sum_{\textbf{k},\alpha}\frac{i\omega_{k}}{c}b_{\textbf{k},\alpha}(t)\textbf{A}_{\textbf{k},\alpha}(\textbf{r})+c.c.,
\end{equation}
\begin{equation}\label{4b}
  b_{\textbf{k},\alpha}(t)=b_{\textbf{k},\alpha}\exp(-i\omega_{k}t),
\end{equation}
\begin{multline}\label{4c}
  \textbf{A}_{\textbf{k},\alpha}(\textbf{r})=
  A_{\textbf{k},\alpha}^{0}\exp(i\textbf{k}_{\shortparallel}\textbf{r}_{\shortparallel})
  \{\textbf{e}_{x}u_{\textbf{k},\alpha}^{x}\sin(k_{n}z)+ \\
  \textbf{e}_{y}u_{\textbf{k},\alpha}^{y}\sin(k_{n}z)+
  \textbf{e}_{z}u_{\textbf{k},\alpha}^{z}\cos(k_{n}z)\}.
\end{multline}
Here $\textbf{e}_{x}$, $\textbf{e}_{y}$ and $\textbf{e}_{z}$ are unit vectors of chosen coordinate system; $\textbf{r}_{\shortparallel}=x\textbf{e}_{x}+y\textbf{e}_{y}$,
$\textbf{k}_{\shortparallel}=k_{x}\textbf{e}_{x}+k_{y}\textbf{e}_{y}$, $k_{n}=\pi n/d$, $n=0,1,2,\ldots$, and
$\textbf{u}_{\textbf{k},\alpha}=\textbf{e}_{x}u_{\textbf{k},\alpha}^{x}+\textbf{e}_{y}u_{\textbf{k},\alpha}^{y}+\textbf{e}_{z}u_{\textbf{k},\alpha}^{z}$ is the unit polarization
vector.

From the equation $div\textbf{E}(\textbf{r},t)=0$ we obtain that $k_{x}u_{\textbf{k},\alpha}^{x}+k_{y}u_{\textbf{k},\alpha}^{y}+ik_{n}u_{\textbf{k},\alpha}^{z}=0$. The modified
polarization vectors $\textbf{u}_{\textbf{k},\alpha}^{\prime}=\textbf{e}_{x}u_{\textbf{k},\alpha}^{x}+\textbf{e}_{y}u_{\textbf{k},\alpha}^{y} +\textbf{e}_{z} i
u_{\textbf{k},\alpha}^{z}$ are orthogonal to the wave vector $\textbf{k}=\textbf{k}_{\shortparallel}+\textbf{e}_{z}k_{n}$ and obeys the following transferability condition
\begin{equation}\label{5}
  \sum_{\alpha}(u_{\textbf{k},\alpha}^{\prime})_{\mu}(u_{\textbf{k},\alpha}^{\prime})_{\nu}^{\ast}=\delta_{\mu\nu}-\frac{k_{\mu}k_{\nu}}{k^{2}}.
\end{equation}
Here $\mu$ and $\nu$ denote the vector projection on the coordinate axes, the sum in (\ref{5}) is  over two orthogonal components of the modified polarization vectors.

To obtain explicit expression of the Schrodinger electric field operator we make the standard replacement
$b_{\textbf{k},\alpha}(t)\rightarrow\sqrt{\hbar/2\omega_{k}}\widehat{a}_{\textbf{k},\alpha}$;
$b_{\textbf{k},\alpha}^{\ast}(t)\rightarrow\sqrt{\hbar/2\omega_{k}}\widehat{a}_{\textbf{k},\alpha}^{\dagger}$, where $\omega_{k}=ck$ is the photon frequency,
$\widehat{a}_{\textbf{k},\alpha}$ and $\widehat{a}_{\textbf{k},\alpha}^{\dagger}$ are the annihilation and creation operators. By this means the electric field operator inside a
Fabry-Perot cavity can be presented as follows:
\begin{multline}\label{6}
  \widehat{\textbf{E}}(\textbf{r})=\sum_{\textbf{k},\alpha}\frac{i\omega_{k}}{c}\sqrt{\frac{\hbar}{2\omega_{k}}}\widehat{a}_{\textbf{k},\alpha}
  A_{\textbf{k},\alpha}^{0} \\
  \{\textbf{e}_{x}u_{\textbf{k},\alpha}^{x}\sin(k_{n}z)+
  \textbf{e}_{y}u_{\textbf{k},\alpha}^{y}\sin(k_{n}z)+ \\
  \textbf{e}_{z}u_{\textbf{k},\alpha}^{z}\cos(k_{n}z)\}\exp(i\textbf{k}_{\shortparallel}\textbf{r}_{\shortparallel})+h.c.
\end{multline}
With this expression we can obtain the magnetic field operator.
\begin{multline}\label{7}
\widehat{\textbf{H}}(\textbf{r})=\sum_{\textbf{k},\alpha}\sqrt{\frac{\hbar}{2\omega_{k}}}\widehat{a}_{\textbf{k},\alpha}A_{\textbf{k},\alpha}^{0}
\{(ik_{y}u_{\textbf{k},\alpha}^{z}-k_{n}u_{\textbf{k},\alpha}^{y}) \\
\cos(k_{n}z)\textbf{e}_{x}+(k_{n}u_{\textbf{k},\alpha}^{x}-ik_{x}u_{\textbf{k},\alpha}^{z}) \\
\cos(k_{n}z)\textbf{e}_{y}+(ik_{x}u_{\textbf{k},\alpha}^{y}-ik_{y}u_{\textbf{k},\alpha}^{x}) \\
\sin(k_{n}z)\}\exp(i\textbf{k}_{\shortparallel}\textbf{r}_{\shortparallel})+h.c.
\end{multline}
Here $A_{\textbf{k},\alpha}^{0}$ is the normalization constant. It can be calculated on the basis of the standard form of the field Hamiltonian $\widehat{H}_{f}$.
\begin{multline}\label{8}
  \widehat{H}_{f}=\int\limits_{V_{q}}\frac{1}{8\pi}(\widehat{\textbf{E}}^{2}+\widehat{\textbf{H}}^{2})dV= \\
  \sum_{\textbf{k},\alpha}\hbar\omega_{k}(\widehat{a}_{\textbf{k},\alpha}^{\dagger}
  \widehat{a}_{\textbf{k},\alpha}+\frac{1}{2}).
\end{multline}
Here $V_{q}$ is the quantization volume, $V_{q}=\{0\leq z\leq d\}\times\{0\leq x,y \leq L\}$. From the equation (\ref{8}) we have
\begin{equation}\label{9}
  A_{\textbf{k},\alpha}^{0}=\sqrt{\frac{8\pi c^{2}}{L^{2}d}}\times
  \begin{cases}
  1, & \text{if $n\in N$} \\
  1/\sqrt{2}, & \text{if $n=0$}.
  \end{cases}
\end{equation}
$L$ is the longitudinal size of the quantization volume.

In accordance with  \cite{27} and \cite{21} we will seek the wave function $\psi$ as an expansion in a set of eigenstates $\{|l\rangle\}$ of the operator $H_{0}$:
\begin{equation}\label{10}
  \psi=\sum_{l}b_{l}(t)|l\rangle.
\end{equation}
Here, the subscript $l$ defines the state of all atoms and the field. Using this representation of the wave function we convert the equation (\ref{1}) to the system of linear
differential equations for the quantum amplitudes
\begin{equation}\label{11}
  i\hbar\frac{\partial b_{l}(t)}{\partial t}-E_{l}b_{l}(t)=\sum_{j}V_{lj}b_{j}(t).
\end{equation}
In this equation $E_{l}$ is the energy of $l$ state of the system, which consists of noninteracting atoms and electromagnetic field.

Because of infinity number of the field states the total number of equations in the system (\ref{11}) is equal to infinity.

The key simplification of the approach employed is in restriction of the total number of states $|l\rangle$ taken into account. We will calculate all radiative correction up to
the second order of the fine structure constant. In this case we can consider only the following states (see \cite{28}):

1. One-fold atomic excited states

$\psi_{e_{a}}=|g,...,g,e,g,...,g\rangle\otimes|vac\rangle$, $E_{e_{a}}=\hbar\omega_{a}$

2. Resonant single-photon states

$\psi_{g}=|g,...,g\rangle\otimes|\textbf{k},\alpha\rangle$, $E_{g}=\hbar\omega_{k}$

3. Nonresonant states with two excited atoms and one photon

$\psi_{e_{a}e_{b}}=|g,...,g,e,g,...,g,e,g,...,g\rangle\otimes|\textbf{k},\alpha\rangle, \\
E_{e_{a}e_{b}}=\hbar(\omega_{a}+\omega_{b})+\hbar\omega_{k}$

In the rotating wave approximation it is enough to take into account only the first and second group of states. Nonresonant states are necessary for a correct description of the
dipole-dipole interaction at short interatomic distances, comparable with $\lambda_{0}$.

For a description of the coherent external light scattering, it is necessary to complete the set of quantum states by the vacuum state without excitation both in atomic and
field subsystem

$\psi_{g\prime}=|g,...,g\rangle\otimes|vac\rangle$, $E_{g\prime}=0$

In the framework of the assumptions considered here, the quantum amplitude of the state $\psi_{g\prime}$ does not change during the evolution of the system. It is explained by
the fact that any transitions between $\psi_{g\prime}$ and the other quantum states taken into account are impossible. The Lamb shift is considered to be included in
$\omega_{0}$.

Despite the restriction of the total number of quantum states, the set of equations remains infinite. We can, however, exclude amplitudes of states with one photon and obtain a
finite closed system of equations for the atomic states $b_{e}$. For Fourier components $b_{e}(\omega)$ we have (at greater length; see \cite{21})
\begin{equation}\label{12}
\sum_{e\prime}\bigl[(\omega-\omega_{a})\delta_{ee'}-\Sigma_{ee'}(\omega)\bigl]b_{e'}(\omega)=i\delta_{e' o}.
\end{equation}

This specific set of equations was obtained under the assumption that at the initial time only one atom is excited. We denote it by the subscript \emph{o}. All other atoms are
in the ground states at $t=0$ and electromagnetic field is in the vacuum state. The system (\ref{12}) with the initial conditions considered here allows us to analyze both
stationary light scattering as well as nonstationary problems (see \cite{21}).

The size of the system (\ref{12}) is determined by the number of atoms $N$ and the structure of theirs energy levels. In this paper we consider the impurity centers with ground
state $J=0$. Total angular momentum of the excited state is $J=1$. It includes three sublevels  $e=|J,m\rangle$, which differ by the value of angular momentum projection on the
quantization axis $m=-1,0,1$. Therefore, the total number of one-fold atomic excited states is $3N$. This scheme of levels corresponds to atoms with 2 valence electrons such as
Sr, Yb, Ca.

The matrix $\Sigma_{ee'}(\omega)$ describes both spontaneous decay and excitation exchange between the atoms.  This matrix can be calculated as follows:
\begin{multline}\label{13}
  \Sigma_{ee'}(\omega)=\sum_{g}V_{e;g}V_{g;e'}\zeta(\hbar\omega-E_{g})+ \\
  \sum_{ee}V_{e;ee}V_{ee;e'}\zeta(\hbar\omega-E_{ee}).
\end{multline}
In this equation $\zeta(x)$ is a singular function which is determined by the relation $\varsigma \left( x\right) =\underset{k\rightarrow \infty }{\lim }(1-\exp (ikx))/x$.

We will calculate the sum over the field variables in the equation (\ref{13}) in the limit $L\rightarrow\infty$. This implies summation over $n$ ($k_n$), the integration over
$k_{\shortparallel}$ and the polar angle $\varphi$ as well as summation over polarization types.
\begin{equation}
\nonumber \sum_{g} \text{or} \sum_{ee}\rightarrow\frac{L^{2}}{(2\pi)^{2}}\sum_{n=0}^{+\infty}{}^\prime\int\limits_{0}^{+\infty}
k_{\shortparallel}dk_{\shortparallel}\int\limits_{0}^{2\pi}d\varphi\sum_{\alpha}.
\end{equation}
The prime sign here denotes an additional coefficient $1/2$ in the sum over $n$ for $n=0$. This coefficient appears from the equation (\ref{9}).

When calculating matrix elements of the operator $\widehat{V}$ in (\ref{13}) we will denote by index $a$ those atoms which transit from excited state to ground one and by index
$b$ atoms which perform reverse transition.   With eqs. (\ref{6}) and (\ref{9}) we have
\begin{multline}\label{14}
V_{e;g}=\langle e|\widehat{V}|g\rangle=-\textbf{d}_{e_{b};g_{b}}i\sqrt{\frac{4\pi\hbar \omega_{k}}{L^{2}d}} \\
\{\textbf{e}_{x}u_{\textbf{k},\alpha}^{x}\sin(k_{n}z_{b})+
  \textbf{e}_{y}u_{\textbf{k},\alpha}^{y}\sin(k_{n}z_{b})+ \\
  \textbf{e}_{z}u_{\textbf{k},\alpha}^{z}\cos(k_{n}z_{b})\}\exp(i\textbf{k}_{\shortparallel}\textbf{r}_{\shortparallel b})
\end{multline}
\begin{multline}\label{15}
V_{g;e'}=\langle g|\widehat{V}|e'\rangle=\textbf{d}_{g_{a};e_{a}}i\sqrt{\frac{4\pi\hbar \omega_{k}}{L^{2}d}} \\
\{\textbf{e}_{x}(u_{\textbf{k},\alpha}^{x})^{\ast}\sin(k_{n}z_{a})+
  \textbf{e}_{y}(u_{\textbf{k},\alpha}^{y})^{\ast}\sin(k_{n}z_{a})+ \\
  \textbf{e}_{z}(u_{\textbf{k},\alpha}^{z})^{\ast}\cos(k_{n}z_{a})\}\exp(-i\textbf{k}_{\shortparallel}\textbf{r}_{\shortparallel a})
\end{multline}
The calculation of $V_{e;ee}$ $V_{ee;e'}$, which can be performed in the same way, gives $V_{e;ee}=V_{e;g}$, $V_{ee;e'}=V_{g;e'}$.

For arbitrary $\omega$, the explicit expression of the matrix $\Sigma_{ee'}(\omega)$ is very complicated. We can, however, simplify it essentially under so-called pole
approximation when its value for frequency $\omega$ is replaced by its value for frequency $\omega_0$ of the atomic resonance. This approximation was studied in detail in
\cite{28a}, where it was shown that it can be applied in systems where retardation effects are insignificant. This condition can be satisfied in the real experiment with a good
accuracy even for dense atomic ensembles. In the pole approximation we get
\begin{multline}\label{16}
\Sigma_{ee\prime}(\omega_{0})=\frac{L^{2}}{4\pi^{2}}\sum_{n=0}^{+\infty}{}^\prime\int\limits_{0}^{+\infty}
k_{\shortparallel}dk_{\shortparallel}\int\limits_{0}^{2\pi}d\varphi \\
\frac{4\pi\hbar\omega_{k}}{L^{2}d}\Biggl\{d_{e_{b};g_{b}}^{x}d_{g_{a};e_{a}}^{x}\sin(k_{n}z_{b})\sin(k_{n}z_{a}) \\
\left(\frac{k_{y}^{2}+k_{n}^{2}}{k^{2}}\right)+d_{e_{b};g_{b}}^{x}d_{g_{a};e_{a}}^{y}\sin(k_{n}z_{b})
\sin(k_{n}z_{a}) \\
\left(-\frac{k_{x}k_{y}}{k^{2}}\right)+d_{e_{b};g_{b}}^{x}d_{g_{a};e_{a}}^{z}\sin(k_{n}z_{b})
\cos(k_{n}z_{a}) \\
\left(-i\frac{k_{x}k_{n}}{k^{2}}\right)+d_{e_{b};g_{b}}^{y}d_{g_{a};e_{a}}^{x}\sin(k_{n}z_{b})\sin(k_{n}z_{a}) \\
\left(-\frac{k_{x}k_{y}}{k^{2}}\right)+d_{e_{b};g_{b}}^{y}d_{g_{a};e_{a}}^{y}\sin(k_{n}z_{b})\sin(k_{n}z_{a}) \\
\left(\frac{k_{x}^{2}+k_{n}^{2}}{k^{2}}\right)+d_{e_{b};g_{b}}^{y}d_{g_{a};e_{a}}^{z}\sin(k_{n}z_{b})\cos(k_{n}z_{a}) \\
\left(-i\frac{k_{y}k_{n}}{k^{2}}\right)+d_{e_{b};g_{b}}^{z}d_{g_{a};e_{a}}^{x}\cos(k_{n}z_{b})\sin(k_{n}z_{a}) \\
\left(i\frac{k_{x}k_{n}}{k^{2}}\right)+d_{e_{b};g_{b}}^{z}d_{g_{a};e_{a}}^{y}\cos(k_{n}z_{b})\sin(k_{n}z_{a}) \\
\left(i\frac{k_{y}k_{n}}{k^{2}}\right)+d_{e_{b};g_{b}}^{z}d_{g_{a};e_{a}}^{z}\cos(k_{n}z_{b})\cos(k_{n}z_{a}) \\
\left(\frac{k_{x}^{2}+k_{y}^{2}}{k^{2}}\right)\Biggl\} \exp(i\textbf{k}_{\shortparallel}\textbf{r}_{\shortparallel ab})
\biggl[-i\pi\delta(\hbar\omega_{0}-\hbar\omega_{k})- \\
i\pi\delta(-\hbar\omega_{0}-\hbar\omega_{k})+
\text{v.p.}\biggl(\frac{1}{\hbar\omega_{0}-\hbar\omega_{k}}+ \\
\frac{1}{-\hbar\omega_{0}-\hbar\omega_{k}}\biggl)\biggl]
\end{multline}
In this equation $\textbf{r}_{\shortparallel ab}=\textbf{r}_{\shortparallel b}-\textbf{r}_{\shortparallel a}$, the sum over polarization types was calculated using the relation
(\ref{5}). The singular $\zeta$-function is represented as follows $\zeta(x)=-i\pi\delta(x)+\text{v.p.}/x$, where v.p. means the principal value of the integral which contains
$\zeta$-function.

The diagonal element of the matrix (\ref{16}) describes the Lamb shift and the natural linewidth of an atom inside a cavity. The dipole approximation used here does not allow us
to calculate the Lamb shift correctly. This manifests itself in the infinity real part of the diagonal element. We can, however, consider that the Lamb shift is included into
$\omega_{0}$. Hereafter we will associate $\omega_{0}$ with the resonant transition frequency taking into account the Lamb shift.

The imaginary part of diagonal element determining the natural linewidth can be calculated as follows
\begin{multline}\label{17}
\Sigma_{ee'}(\omega_{0})\Bigl|_{e=e'}=-\frac{i\pi}{d}d_{e_{a};g_{a}}^{z}d_{g_{a};e_{a}}^{z}\frac{\omega_{0}^{2}}{c^{2}}- \\
\frac{i\pi}{d}
\sum_{n=1}^{\left[\left[\frac{\omega_{0}d}{\pi c}\right]\right]}\biggl\{\left( \frac{\omega_{0}^{2}}{c^{2}}+k_{n}^{2}\right)\sin^{2}(k_{n}z_{a}) \\
\Bigl(d_{e_{a};g_{a}}^{x}d_{g_{a};e_{a}}^{x}+ d_{e_{a};g_{a}}^{y}d_{g_{a};e_{a}}^{y}
\Bigl)+ \\
2\left(\frac{\omega_{0}^{2}}{c^{2}}-k_{n}^{2}\right)\cos^{2}(k_{n}z_{a}) d_{e_{a};g_{a}}^{z}d_{g_{a};e_{a}}^{z}\biggl\}.
\end{multline}
Double brackets here means the integer part.

If $e'$ and $e$ correspond to excited states of different atoms, for example atoms $a$ and $b$, matrix element $\Sigma_{ee'}(\omega)$ describes excitation exchange between these
atoms. As it is known this exchange is responsible for interatomic dipole-dipole interaction. Matrix element $\Sigma_{ee'}(\omega)$ is easy to calculate in the coordinate frame
with X-axis along the vector $\textbf{r}_{\shortparallel ab}$. In this frame $k_{x}=k_{\shortparallel}\cos{\varphi}$, $k_{y}=k_{\shortparallel}\sin{\varphi}$ and
$\textbf{k}_{\shortparallel}\textbf{r}_{\shortparallel ab}=k_{\shortparallel}r_{\shortparallel ab}\cos{\varphi}$. Double integral in the eq. (\ref{16}) can be simplified by the
following relations
\begin{equation}
\nonumber k_{x}\exp(i\textbf{k}_{\shortparallel}\textbf{r}_{\shortparallel ab})=-i\frac{\partial}{\partial x_{ab}}\exp(i\textbf{k}_{\shortparallel}\textbf{r}_{\shortparallel
ab}),
\end{equation}
\begin{equation}
\nonumber k_{x}^{2}\exp(i\textbf{k}_{\shortparallel}\textbf{r}_{\shortparallel ab})=-\frac{\partial^{2}}{\partial
x_{ab}^{2}}\exp(i\textbf{k}_{\shortparallel}\textbf{r}_{\shortparallel ab}),
\end{equation}
\begin{equation}
\nonumber k_{x}k_{y}\exp(i\textbf{k}_{\shortparallel}\textbf{r}_{\shortparallel ab})=-\frac{\partial^{2}}{\partial
x_{ab}y_{ab}}\exp(i\textbf{k}_{\shortparallel}\textbf{r}_{\shortparallel ab}).
\end{equation}
Here $x_{ab}=x_{b}-x_{a}$, $y_{ab}=y_{b}-y_{a}$. For the other items in (\ref{16}) we have similar relations, and we have
\begin{multline}\label{18}
\Sigma_{ee'}(\omega_{0})\Bigl|_{a\neq b}=\sum\limits_{n=0}^{+\infty}{}^\prime\widehat{A}_{n}\int\limits_{0}^{+\infty}
k_{\shortparallel}dk_{\shortparallel}\int\limits_{0}^{2\pi}d\varphi\frac{c}{\pi d}\frac{1}{k} \\
\exp(ik_{\shortparallel}r_{\shortparallel ab}\cos{\varphi})\biggl[-i\pi\delta(\omega_{0}-c k)+ \\ \frac{2c k}{\omega_{0}^{2}-c^{2}k^{2}}\biggl]
\end{multline}
The differential operator $\widehat{A}_{n}$ is determined as follows
\begin{multline}\label{19}
\widehat{A}_{n}=d_{e_{b};g_{b}}^{x}d_{g_{a};e_{a}}^{y}\sin(k_{n}z_{b})\sin(k_{n}z_{a})
\left(\frac{\partial^{2}}{\partial x \partial y}\right)\\
+d_{e_{b};g_{b}}^{x}d_{g_{a};e_{a}}^{x}\sin(k_{n}z_{b})\sin(k_{n}z_{a})
\left(k_{n}^{2}-\frac{\partial^{2}}{\partial y^{2}}\right)\\
+d_{e_{b};g_{b}}^{x}d_{g_{a};e_{a}}^{z}\sin(k_{n}z_{b})\cos(k_{n}z_{a}) \left(-k_{n}\frac{\partial}{\partial x}\right) \\+
d_{e_{b};g_{b}}^{y}d_{g_{a};e_{a}}^{x}\sin(k_{n}z_{b})\sin(k_{n}z_{a}) \left(\frac{\partial^{2}}{\partial x \partial y}\right)\\+
d_{e_{b};g_{b}}^{y}d_{g_{a};e_{a}}^{y}\sin(k_{n}z_{b})\sin(k_{n}z_{a}) \left(-\frac{\partial^{2}}{\partial x^{2}}+k_{n}^{2}\right)\\+
d_{e_{b};g_{b}}^{y}d_{g_{a};e_{a}}^{z}\sin(k_{n}z_{b})\cos(k_{n}z_{a}) \left(-k_{n}\frac{\partial}{\partial y}\right)\\+
d_{e_{b};g_{b}}^{z}d_{g_{a};e_{a}}^{x}\cos(k_{n}z_{b})\sin(k_{n}z_{a}) \left(k_{n}\frac{\partial}{\partial x}\right)\\-
d_{e_{b};g_{b}}^{z}d_{g_{a};e_{a}}^{z}\cos(k_{n}z_{b})\cos(k_{n}z_{a}) \left(\frac{\partial^{2}}{\partial x^{2}}+\frac{\partial^{2}}{\partial y^{2}}\right)\\+
d_{e_{b};g_{b}}^{z}d_{g_{a};e_{a}}^{y}\cos(k_{n}z_{b})\sin(k_{n}z_{a}) \left(k_{n}\frac{\partial}{\partial y}\right).
\end{multline}
Here $x=x_{b}-x_{a}$, $y=y_{b}-y_{a}$.

The calculation of the double integral in eq. (\ref{18}) produces Bessel functions $J_{0}$, $K_{0}$ and $N_{0}$
\begin{multline}\label{20}
\Sigma_{ee'}(\omega_{0})\Bigl|_{a\neq b}=\frac{2\pi}{d}\sum\limits_{n=0}^{\left[\left[\frac{\omega_{0}d}{\pi c}\right]\right]}{}^\prime\widehat{A}_{n} \\
\Biggl[N_{0}\left(r_{\shortparallel ab}\sqrt{\frac{\omega_{0}^{2}}{c^{2}}-k_{n}^{2}}\right) - iJ_{0}\left(r_{\shortparallel
ab}\sqrt{\frac{\omega_{0}^{2}}{c^{2}}-k_{n}^{2}}\right)\Biggl] \\ -\frac{4}{d}\sum\limits_{n=\left[\left[\frac{\omega_{0}d}{\pi c}\right]\right]+1}^{+\infty}\widehat{A}_{n}
K_{0}\left(r_{\shortparallel ab}\sqrt{k_{n}^{2}-\frac{\omega_{0}^{2}}{c^{2}}}\right)
\end{multline}
The differential operator $\widehat{A}_{n}$ yields  bulky expressions which we do not show here.

The equations (\ref{17}) and (\ref{20}) for the matrix $\Sigma_{ee'}(\omega_{0})$ obtained in this section allows us to solve the set of equations (\ref{12}) numerically and
obtain, on this background, the Fourier-amplitudes of atomic states $b_{e}(\omega)$. Using $b_{e}(\omega)$ we can obtain the amplitudes of all states taken into account in our
calculations (see \cite{21}) and, consequently, the wave function of the considered system.

In the next section, we will use the obtained general equations to calculate the transition spectrum of an excited atom and the time dependence of the total excitation of atomic
ensemble. On this basis, we will analyze radiation trapping in considered system.
\section{Results and discussion}
The influence of the dipole-dipole interaction on the properties of atomic ensemble is determined not only by the atomic density. The shifts of the energy levels caused by the
internal fields of a dielectric are also very important. The value of these shifts depends on a number of factors, first of all, on the nature of the dielectric and its
temperature. As it was mentioned above in the present paper we will assume that the temperature is low enough to neglect the electron-phonon interaction. So the spectral lines
of impurity centers are Zero-phonon.

The shift of the transition line can be presented as a sum of its average value $\overline{\Delta}$ and some random contribution connected with the inhomogeneity of the internal
fields of the dielectric. We consider this random contribution to be normally distributed with r.m.s. deviation $\delta$. The ratio of $\delta$ to natural line width of the
atoms $\gamma_{0}$ characterizes the degree of resonance between impurity atoms. This is one of a key parameters of the considered system in our theory.

Depending on the symmetry of the internal fields of a dielectric, the average shift $\overline{\Delta}$ can be both the same for all Zeeman sublevels of excited state and
different. In general, the theory allows us to analyze both cases. Only for specifics, in this section we limit ourselves by the first case. This corresponds to the cubic
symmetry of internal fields of a medium, for instance. Hereafter we will consider $\overline{\Delta}$ to be included in the resonant transition frequency $\omega_{0}$.

In the framework of the general approach we can consider an arbitrary distance between the mirrors of a Fabry-Perot cavity. However, the most exciting case is $d<\lambda_{0}/2$
due to the suppression of the spontaneous decay of the states $m=\pm1$ in the cavity. So we focus our attention on this case. Taking $k_{0}^{-1}$ as a unit of length, hereafter
we consider $d=3$.

In this paper we assume spatially localized initial excitation of the ensemble. Such initial condition can be prepared by a two-photon resonance method. In the framework of this
method the sample is illuminated by two narrow and off-resonant orthogonally propagated light beams (both beams parallel to the mirrors of a cavity). Each beam does not cause
single-photon excitation, but their simultaneous interaction with atoms in the crossing region cause two-photon excitation from the ground $S$ to the high-energy excited $D$
state if conditions of two-photon resonance are satisfied. If the transition frequency from the high-energy $D$ state to the studied $P$ state is high enough so that its
resonant wavelength $\lambda_{D\rightarrow P}<2d$, this spontaneous transition leads to population of $P$ state. Note that the cascade transition from $D$ to $S$ state can be
forbidden, in particular, due to the spontaneous decay suppression in each step of the cascade. So the registration of the photon resonant to the transition $D\longrightarrow P$
means the $P$ state population.

The thereby described method allows obtaining small cluster of excited atoms in the middle of the sample. For simplicity thereafter in the paper we will consider that at initial
time only one atom located in the center of a sample is excited. Note that possibilities of two-photon excitation $5s$ $S$ -- $2(1/2)$ $\longrightarrow$ $5p$ $P$ -- $2(j)$
$\longrightarrow$ $5d$ $D$ -- $2(j)'$ of rubidium atoms have been already studied in Ref. \cite{Hav}.

Eq. (\ref{17}) shows that the natural linewidth of the excited atom inside the cavity depends on its $z$ position even in the case of a single atom. So all the results must
depend on this parameter. In the framework of the general theory we can consider an arbitrary position of all the atoms, including the excited initially atom. From the
experimental point of view the position of excited atom is determined by the crossing region of two beams. Further we will consider $z_{exc}=d/2$.

Note that the matrix $\Sigma_{ee'}$ and subsequently any physical observable depends on the positions of all impurity atoms. In this paper we analyze spatially disordered atomic
ensembles with uniform (on average) distribution of atomic density. So we average all the results over random spatial configurations of the ensemble as well as over random
shifts of energy levels caused by the inhomogeneity of the internal fields of a dielectric. This averaging is performed by a Monte Carlo method.
\subsection{Atomic transition spectrum}
Figure 1 shows the transition spectrum of the central atom, which is initially excited. The calculations were made for $\delta=0$. In this case all the atoms are resonant to
each other, so the role of the dipole-dipole interaction is maximal. Ensembles with two densities $n=0.01$ and $n=0.05$ are considered.

\begin{figure}[th]
\begin{center}
{$\scalebox{0.4}{\includegraphics*{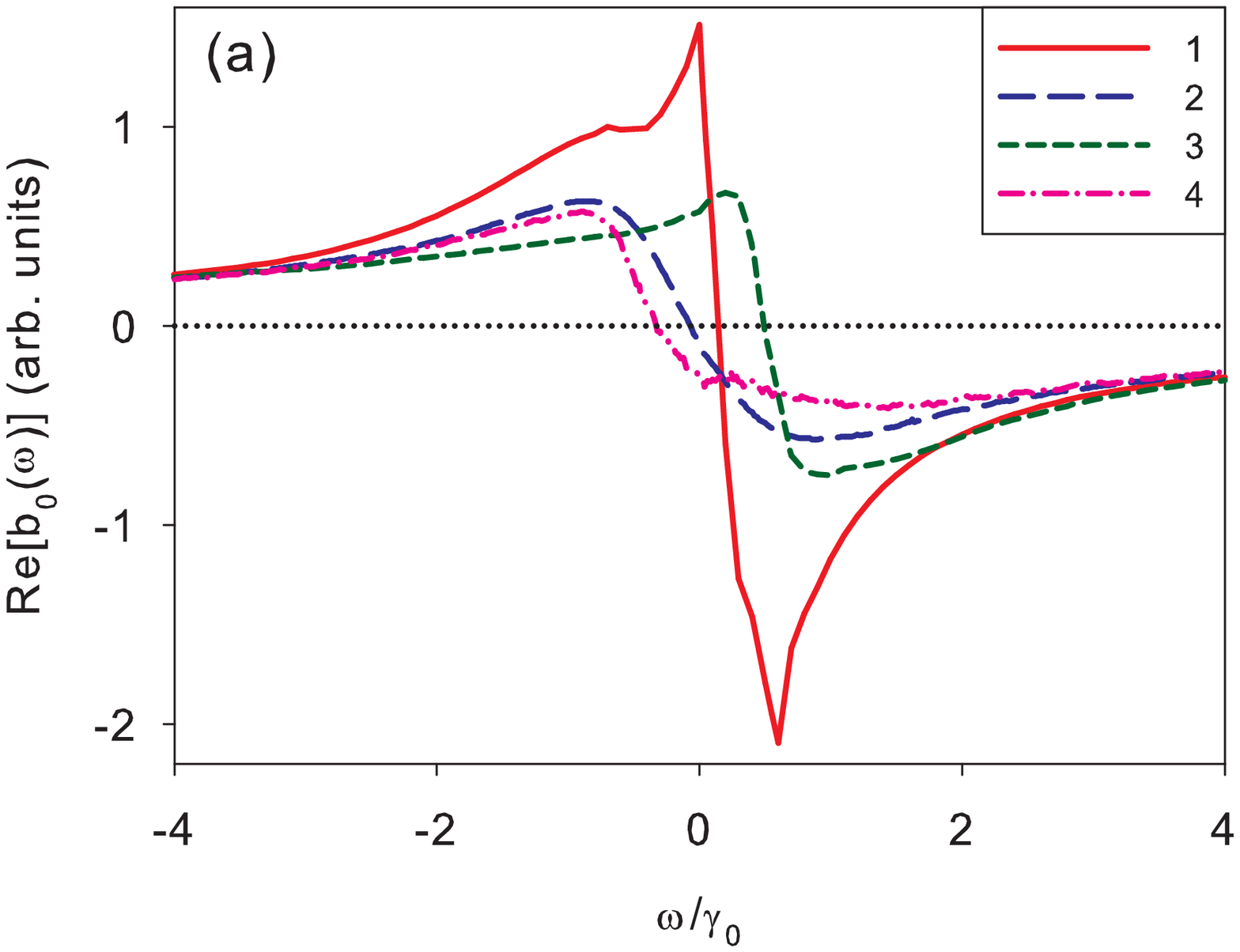}}$ }{$\scalebox{0.4}{%
\includegraphics*{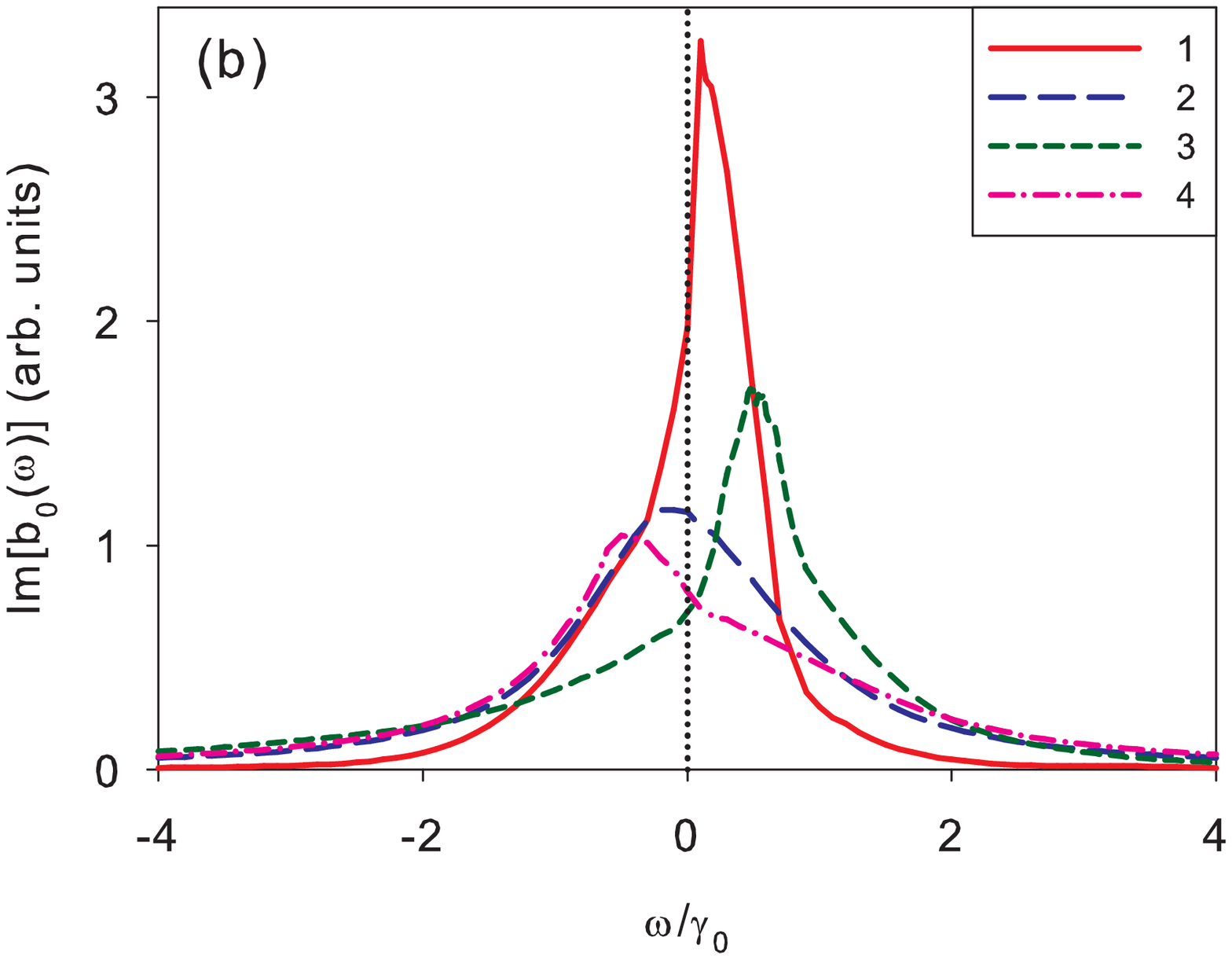}}$ }
\caption{Transition spectrum of an atom inside a microcavity, $d=3$, $\delta=0$, real part (a), imaginary part (b), 1 -- $n=0.01$, $m=\pm1$, 2 -- $n=0.01$, $m=0$, 3 -- $n=0.05$,
$m=\pm1$, 4 -- $n=0.05$, $m=0$.}
\end{center}
\par
\label{fig1}
\end{figure}

Density $n=0.01$ is small and in this case cooperative effects in the free space manifest themselves slightly, see \cite{29} for detail. Difference between transition spectrum
of the free atom and atom excited  in the ensemble when the cavity is absent is very small. However, in a cavity the dipole-dipole interaction transforms the transition spectrum
significantly. First of all, in the Figure 1 we see that the transition spectrum in a cavity for Zeeman sublevels $m=\pm1$ extremely differs from one for $m=0$. Despite the
strong cavity suppression of the spontaneous decay from sublevels $m=\pm1$ \cite{2}, we observe that in the ensemble of the density $n=0.01$ the width of the transition spectrum
is approximately equal to $0.6\gamma_{0}$. This broadening is determined completely by polyatomic cooperative effect. Also we observe some blue shift as well as an essential
discrepancy between the spectrum shape and a typical Lorentz profile. For $m=0$ the modification of the spectrum shape is considerably less than for $m=\pm1$. It can be
explained by the fact that for $m=0$ the spontaneous decay of a single atom in a cavity is not suppressed unlike sublevels $m=\pm1$. However, a noticeable red shift is observed
even in this case.

In the case of essentially higher density $n=0.05$ the dipole-dipole interaction plays an important role for the atomic ensembles in free space, without cavity \cite{17}.
Nevertheless, the Fig.1 shows that microcavity  modifies the transition spectrum additionally. In the Fig.1 one can see a blue shift comparable with $\gamma_{0}$ for $m=\pm1$
and approximately the same red one for $m=0$. The shape of the transition spectrum significantly differs from a Lorentz profile for any Zeeman sublevel due to an essential role
of the the dipole-dipole interaction. This interaction causes also broadening of the spectrum as density increases.

\subsection{Time dependence of the total excitation of atomic system}
The inverse Fourier transform of $b_{e}(\omega)$ allows us to obtain the time dependence of the quantum amplitudes of the one-fold atomic excited states.
\begin{equation}\label{21}
b_{e}(t)=\int\limits_{-\infty}^{+\infty}\frac{id\omega}{2\pi}\exp(-i\omega t)R_{eo}(\omega).
\end{equation}
Here matrix  $R_{eo}(\omega)$ is the resolvent of the considered system projected on the one-fold atomic excited states \cite{21}. It is determined from the eq. (\ref{12}) as
follows.
\begin{equation}\label{22}
R_{ee'}(\omega)=\bigl[(\omega-\omega_{a})\delta_{ee'}-\Sigma_{ee'}(\omega_{0})\bigl]^{-1}.
\end{equation}

The total excited state population $P_{sum}(t)$ is given by a sum of $|b_{e}(t)|^{2}$ over all atoms in the ensemble. Besides $P_{sum}(t)$ we calculate the time-dependent
collective decay rate:
\begin{equation}\label{23}
\gamma(t)=-\frac{1}{P_{sum}(t)}\frac{d P_{sum}(t)}{dt}.
\end{equation}

Figure 2 shows the time dependence of the total excited state population and the collective decay rate in the case $\delta=0$ both for the ensemble in free space and in the
cavity. The results are presented for the atomic density $n=0.1$ and for the size a sample $R=14$. $R$ means the radius of the spherical sample in the case of free space and the
radius of a cylindrical sample in the case of a cavity. First of all, we observe that the total excited state population in the case of a microcavity decreases slower than one
in the case of free space. Besides that, the decay rate of sublevels $m=\pm1$ in a cavity is less than of the Zeeman sublevel $m=0$. It is connected with mentioned features of
the field modes structure in the microcavity.

\begin{figure}[th]
\begin{center}
{$\scalebox{0.4}{\includegraphics*{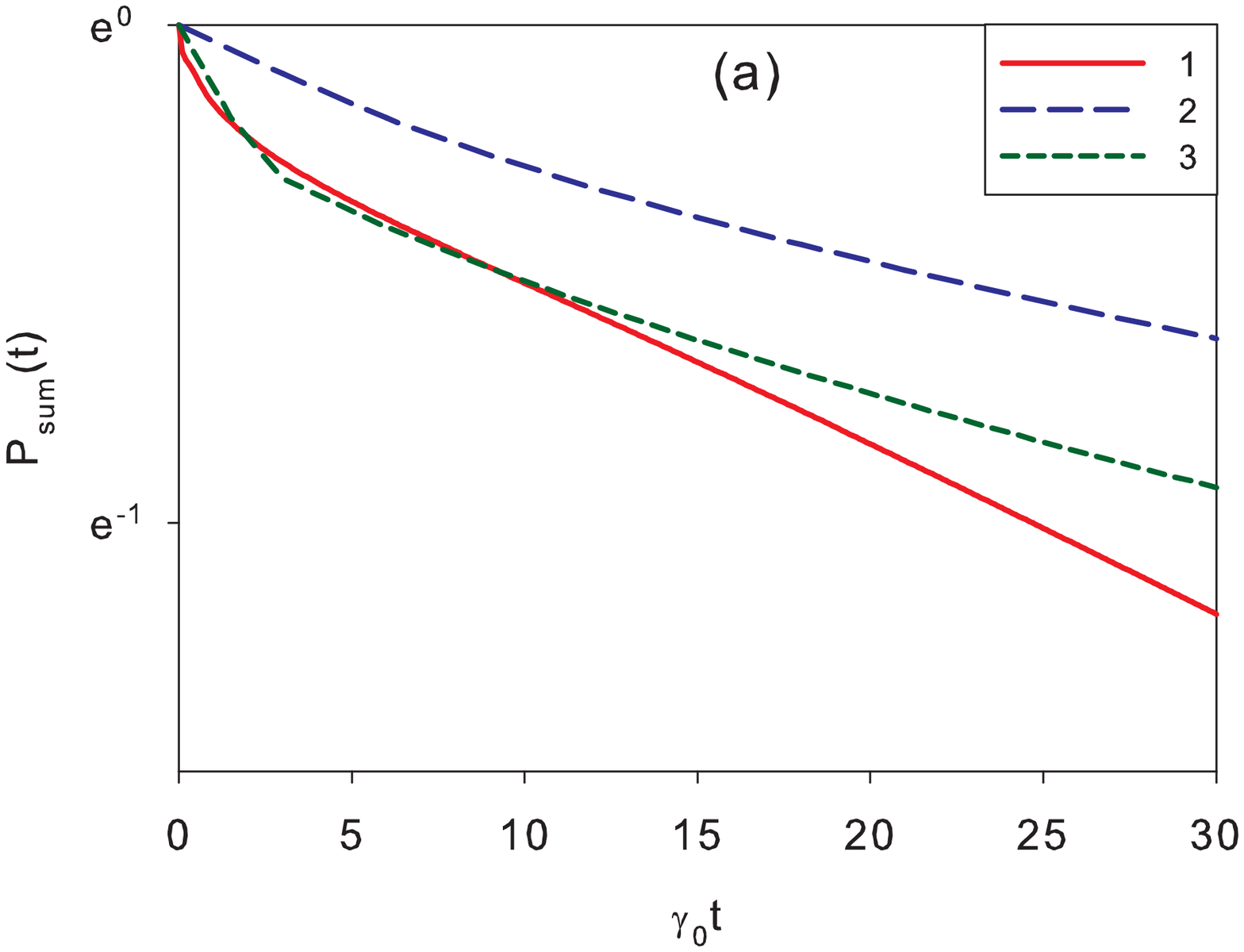}}$ }{$\scalebox{0.4}{%
\includegraphics*{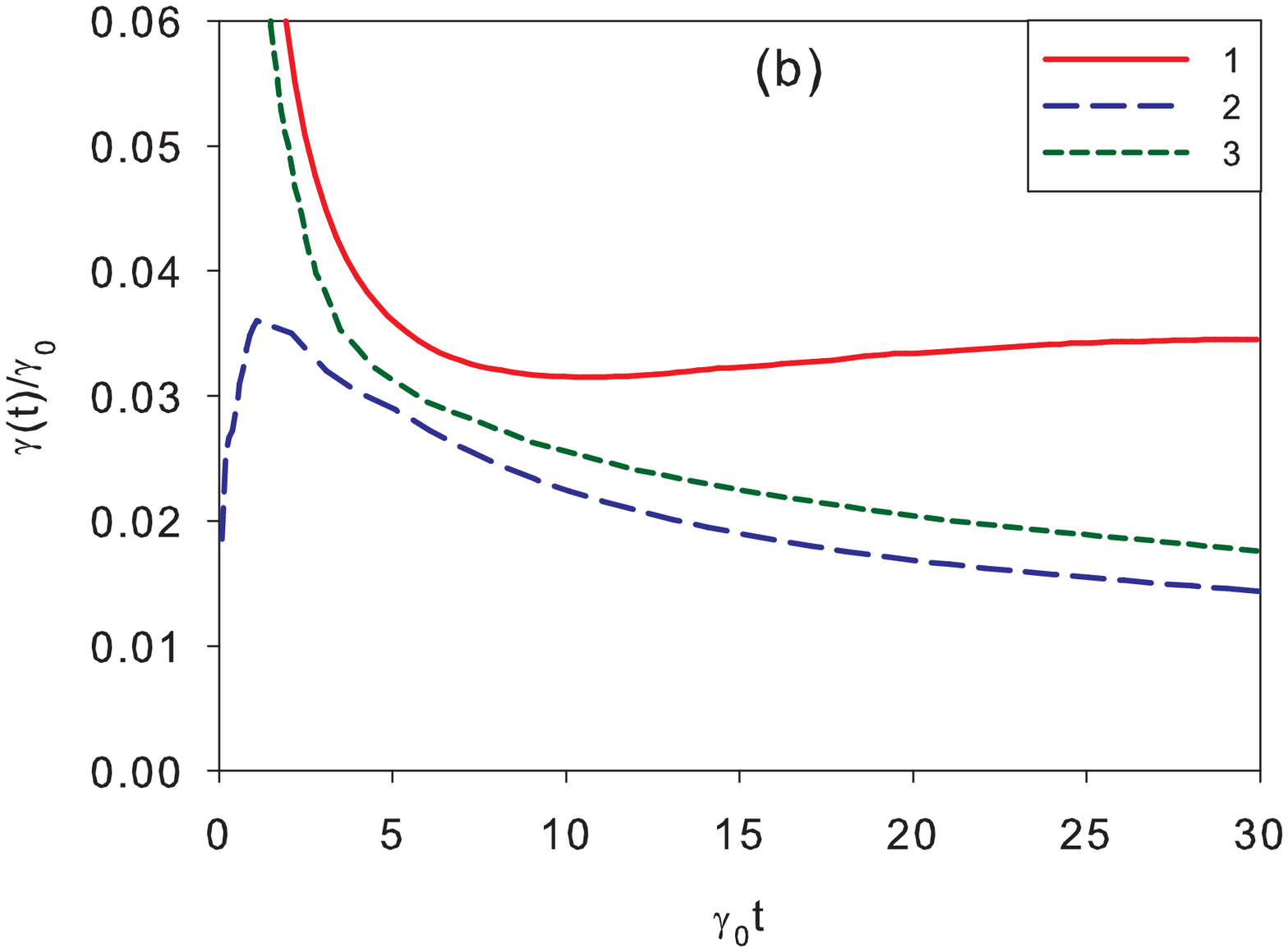}}$}
\caption{Time dependence of the total excited state population (a), collective decay rate (b), $n=0.1$, $\delta=0$, $R=14$; 1 -- free space; 2 -- microcavity, $d=3$, $m=\pm1$; 3
-- microcavity, $d=3$, $m=0$.}
\end{center}
\par
\label{fig2}
\end{figure}

For the time interval $t\gg\gamma_{0}^{-1}$ the time dependence of the total excited state population in the semi-logarithmic scale is close to linear, and subsequently the
collective decay rate depends on time weakly. This case  is similar to a Holstein mode decay. For $t\sim\gamma_{0}^{-1}$ the time dependence of the total excited state
population is more complex because both superradiant and subradiant collective states influence on it. For the atomic ensemble in free space we observe that the collective decay
rate decreases with time here. It can be explained by the fact that the influence of superradiant states decreases with time whereas the influence of subradiant states increases
\cite{17}. The same holds true for the microcavity in the case of sublevel $m=0$ decay. However, the time dependence of the collective decay rate of Zeeman sublevels $m=\pm1$ in
the microcavity is not a monotonic function, and it has a maximum at $t=1.2\gamma_{0}^{-1}$ for considered parameters. To understand this effect we studied the spectral
distribution of the density of collective states. The performed analysis shown that the frequency distribution of the density of states with proper lifetimes has two peaks. It
causes the quantum beats which manifest themselves in the excited state population. Generally, quantum beats influence on the collective decay rate both for $m=\pm1$ and $m=0$.
However, for $m=\pm1$ the role of the described mechanism is more significant, which is connected with the suppression of the spontaneous decay of a single atom in the
microcavity.
\subsection{The time of radiation trapping}
We will estimate the typical time of radiation trapping $\tau$ from the relation $P_{sum}(\tau)=1/e$. In this section we will concentrate our attention on the case of Zeeman
sublevels $m=\pm1$ initial excitation in view of the fact that it provides radiation trapping longer than that corresponding to $m=0$. In addition, we point out that the time of
radiation trapping in the microcavity usually bigger than in the case of the atomic ensemble with the same density in free space. For example, in the case $n=0.1$, $R=14$,
$\delta=0$ we have $\tau=60\tau_{0}$ in a cavity ($m=\pm1$), whereas $\tau=25\tau_{0}$ in a free space. Here $\tau_0=1/\gamma_0$ is natural lifetime of excited states of the
free atom.

Fig.3a shows $\tau$ depending on the size of a sample. The atomic density is chosen $n=0.1$. The results are presented for different values of the r.m.s. deviation of the
inhomogeneous shifts of the resonant transition frequency. The dependence $\tau(R)$ is complex but as the size of the system increases it approaches to parabola. In the case of
mutually resonance impurities $\delta=0$ we observe quadratic dependence $\tau/\tau_0\propto R^{2}$ with a good accuracy starting approximately with $R\approx 15$. It
corresponds to the case when the size of the system is much greater than photon mean free path $l_{ph}$. The latter can be estimated on the basis of calculation
\cite{18}-\cite{19} as $l_{ph}\sim1.6$. Observed quadratic dependence is typical for diffuse radiation transfer.

\begin{figure}[th]
\begin{center}
{$\scalebox{0.4}{\includegraphics*{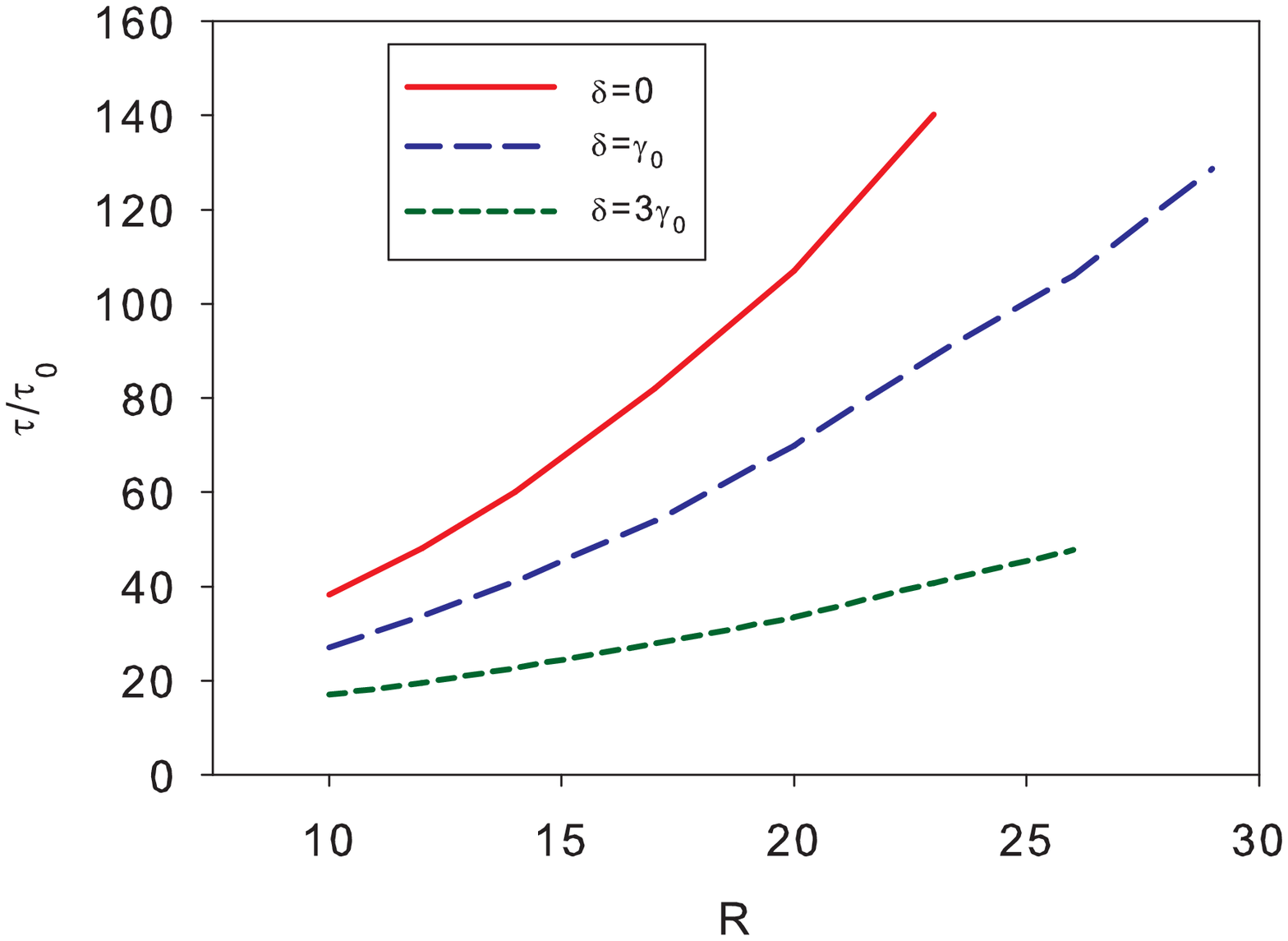}}$ }{$\scalebox{0.4}{%
\includegraphics*{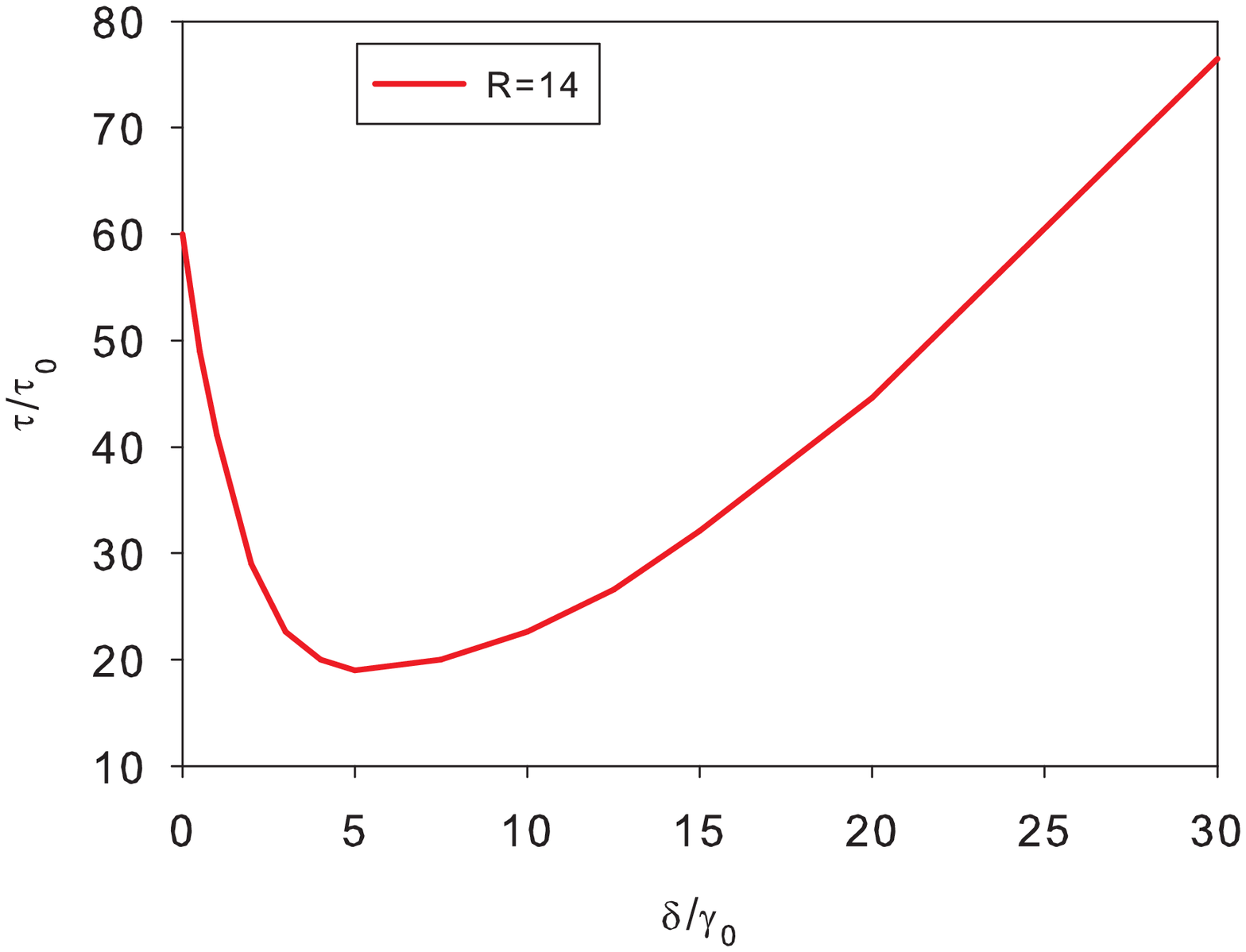}}$}
\caption{The time of radiation trapping in a microcavity, $n=0.1$, $d=3$, $m=\pm1$.}
\end{center}
\par
\label{fig3}
\end{figure}

As $\delta$ increases, the mean free path of photon also increases.  In the case $\delta\neq0$ the regime close to diffuse radiation transfer is achieved for bigger systems. The
dependence $\tau(\delta)$ for $R=14$ is shown in the Fig. 3b. The character of this dependence is determined by two different factors. On the one hand, radiation trapping is
connected with the cooperative multiple scattering. The influence of this mechanism decreases with increasing in $\delta$. On the other hand, trapping time depends on the
spontaneous decay suppression, which manifest itself more noticeably as $\delta$ increases. For small $\delta$ the first mechanism is more significant, so $\tau(\delta)$
decreases. In the case of large $\delta$ decay suppression dominates, this leads to increasing in $\tau$. Generally, the dependence $\tau(\delta)$ has a minimum.

\begin{center}
\emph{Nonresonant impurity centers} ($\delta\gg\gamma_{0}$)
\end{center}

In a range of solid dielectrics the shifts of resonant transition frequency of impurity centers $\delta$ are relatively large. For instance it is typical for NV-centers in a
diamond. If $\delta\gg\gamma_{0}$ the average cross section associated with individual atom is much less than $\lambda_{0}^{2}$. In such a case the dipole-dipole interaction can
be significant only for high density of impurities, when the average distance between mutually resonant atoms  $(n\gamma_{0}/\delta)^{-1/3}$ is less or comparable with
wavelength $\lambda_{0}$ or to put it differently when mean free path of photon $l_{ph}=(n\sigma_{0}\gamma_{0}/\delta)^{-1}$ satisfies the inequality $l_{ph}\leq\lambda_{0}$.
Here $\sigma_{0}=3\lambda_{0}^{2}/2\pi$ is the resonant cross section concerning to free atom.

Assuming the random inhomogeneous shifts of the transition frequency to be normally distributed, we have
\begin{equation}\label{24}
\frac{dn}{d\Delta}=\frac{n_{f}}{\delta\sqrt{2\pi}}\exp\left(-\frac{\Delta^{2}}{2\delta^{2}}\right).
\end{equation}
In this equation $n_{f}$ means the total density of impurity atoms.

In considered case $\delta\gg\gamma_{0}$ not all the atoms in the ensemble essentially influence on the radiative processes but only those  which have the inhomogeneous shifts
close to that of initially excited atom. We will denote the latter as $\Delta_{e}$. In our calculation we take into consideration only atoms with inhomogeneous shifts
$\Delta\in[\Delta_{e}-\Delta_{1};\Delta_{e}+\Delta_{1}]$, where $\Delta_{1}$ is some computational cut-off frequency. We choose  $\Delta_{1}$ so big that the obtained results do
not change with further increasing of this parameter.

Fig. 4 shows the dependence of the time of radiation trapping on the size of a sample for two different densities $n_{f}=5$ and $n_{f}=2$. The r.m.s. deviation of the
inhomogeneous shifts is $\delta=10^{3}\gamma_{0}$.  The mean free path of a photon can be estimated as $l_{ph}=10.6$ and $l_{ph}=26.5$ for $n_{f}=5$ and $n_{f}=2$ respectively.
In these cases the investigation of the dependence $\tau(R)$ in the regime of diffuse radiation transfer $R\gg l_{ph}$ is connected with essential computational difficulties. We
limit ourselves by the case $R\geq l_{ph}$.

\begin{figure}[th]
\begin{center}
{$\scalebox{0.4}{%
\includegraphics*{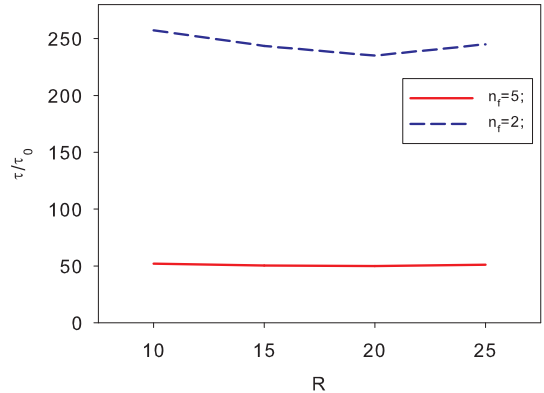}}$}
\caption{The time of radiation trapping in a microcavity in the case of nonresonant impurity centers, $\delta=10^{3}\gamma_{0}$ $d=3$, $m=\pm1$.}
\end{center}
\par
\label{fig4}
\end{figure}

Fig. 4 demonstrates very weak dependence of the time of radiation trapping on the size of the sample. It is explained by competition of two effects. The first one is spontaneous
decay suppression and the second is trapping under multiple scattering. The role of multiple scattering increases with size, whereas the spontaneous decay is substantially
suppressed for small $R$. This suppression explains also increasing of trapping time with density decreasing. As we approach to single atom limit the time of radiation trapping
increases up to infinity.

\section{Conclusion}

We have developed a consistent quantum mechanical theory of cooperative effects in ensembles of point-like impurity centers imbedded into transparent dielectric and located into
Fabry-Perot cavity. Our approach is based on solution of non steady-state Schrodinger equation for the wave function of the joint system consisting of ensemble of motionless
atoms  and the electromagnetic field. The interaction of impurity atoms with the dielectric is simulated by introduction inhomogeneous shifts of the atomic energy levels. The
general approach  allows us to analyze atomic ensembles with arbitrary shape and spatial distribution of impurities.

As an example of a practical implementation of this approach in the present work we study the simultaneous influence of the cavity and resonant dipole-dipole interaction on the
shape of the line of atomic transition as well as on light trapping in dense impurity ensembles. We analyze this influence depending on the size of the ensembles, its density,
as well as the on r.m.s. deviation of the transition frequency shifts caused by the symmetry disturbance of the internal fields of the dielectric medium. The special attention
is given to the case when the distance between two mirrors is less than a half of the transition wavelength. This case is of particular interest due to practically complete
suppression of spontaneous decay of some Zeeman sublevels of atomic exited state.

In our opinion, the theory described in the present paper can be further used for the investigation of Anderson localization of light in the ensembles of impurity centers. It
can be done on the basis of the spectral analysis of the collective states in such ensembles \cite{30} -- \cite{31}. A special attention here should be paid to the case when the
average shift of the transition frequency of impurity atoms caused by the internal fields of a dielectric $\overline{\Delta}$ is different for different Zeeman sublevels. In
this case the excited state is not degenerate, which promotes the Anderson localization \cite{31}.

The developed theory can be generalized to the case of the real susceptibility of the dielectric. In addition, it can be further generalized to the atomic ensembles in the
waveguide. The case when the resonant frequency of atomic transition is less than the cut-off frequency of the waveguide attracts particular interest due to spontaneous decay
suppression of all the Zeeman sublevels. Moreover, the analysis of the atomic systems in a waveguide can be useful for the investigation of Anderson localization, because in
quasi-1D systems all the collective states are localized \cite{32} -- \cite{33}.

\section*{Acknowledgments}

We acknowledge financial support from  the Russian Foundation for Basic Research (Grant No. RFBR-15-02-01013) and from the Ministry of Education and Science of the Russian
Federation (State Assignment 3.1446.2014K). A.S.K. also thanks State Assignment 2014/184, RFBR-16-32-00587, the Council for Grants of the President of the Russian Federation,
and the nonprofit foundation “Dynasty.”

\end{document}